\documentclass[12pt,preprint]{aastex}
\def\amin{\ifmmode^{\prime}\else$^{\prime}$\fi}
\def\asec{\ifmmode^{\prime\prime}\else$^{\prime\prime}$\fi}
\def\ROSAT{\it ROSAT}

\begin{document}

\title{Candidate Isolated Neutron Stars and Other Optically Blank X-ray Fields Identified from the $\ROSAT$ All-Sky and Sloan Digital Sky Surveys\altaffilmark{1}}

\author{
Marcel A. Ag\"ueros\altaffilmark{2,3},
Scott F. Anderson\altaffilmark{2},
Bruce Margon\altaffilmark{4},
Bettina Posselt\altaffilmark{5},
Frank Haberl\altaffilmark{5},
Wolfgang Voges\altaffilmark{5},
James Annis\altaffilmark{6}, 
Donald P. Schneider\altaffilmark{7},
Jonathan Brinkmann\altaffilmark{8}} 

\altaffiltext{1}{Includes observations obtained with the Apache Point Observatory 3.5-m telescope, which is owned and operated by the Astrophysical Research Consortium.}
\altaffiltext{2}{Department of Astronomy, University of Washington, Box 351580, Seattle, WA 98195; agueros@astro.washington.edu}
\altaffiltext{3}{NASA Harriett G. Jenkins Fellow} 
\altaffiltext{4}{Space Telescope Science Institute, 3700 San Martin Drive, Baltimore, MD 21218}
\altaffiltext{5}{Max-Planck-Institut f\"ur extraterrestrische Physik, Geissenbachstrasse 1, D-85741 Garching, Germany} 
\altaffiltext{6}{Fermi National Accelerator Laboratory, P.O. Box 500, Batavia, IL 60510} 
\altaffiltext{7}{Department of Astronomy and Astrophysics, 525 Davey Laboratory, Pennsylvania State University, University Park, PA 16802}
\altaffiltext{8}{Apache Point Observatory, P.O. Box 59, Sunspot, NM 88349}

\begin{abstract}
Only seven radio-quiet isolated neutron stars (INSs) emitting thermal
X rays are known, a sample that has yet to definitively address 
such fundamental issues as the equation of state of
degenerate neutron matter. We describe a selection algorithm 
based on a cross-correlation of the $\ROSAT$ All-Sky Survey 
(RASS) and the Sloan Digital Sky Survey (SDSS) that 
identifies X-ray error circles devoid of plausible optical counterparts 
to the SDSS $g \sim 22$ magnitudes limit. We quantitatively characterize
these error circles as optically blank; they may host INSs or other similarly
exotic X-ray sources such as radio-quiet BL~Lacs, obscured AGN, etc.
Our search is an order of magnitude more selective than 
previous searches for optically blank RASS error circles, and excludes
the $99.9\%$ of error circles that contain more common X-ray-emitting 
subclasses. We find $11$ candidates, nine of which are new. 
While our search is designed to find the best INS candidates and not to 
produce a complete list of INSs in the RASS, it is reassuring that our
number of candidates is consistent with predictions from INS population models. 
Further X-ray observations will obtain pinpoint positions 
and determine whether these sources are entirely optically blank at $g \sim 22$, 
supporting the presence of likely isolated neutron stars 
and perhaps enabling detailed follow-up studies of neutron star physics.

\end{abstract}


\section{                         Introduction                    }
Neutron stars were empirically confirmed first as radio pulsars \citep[][]{hewish},
and these objects continue to dominate neutron star (NS) statistics. Currently there 
are over $1500$ pulsars cataloged, and the number grows steadily\footnote{For an up-to-date catalog of known pulsars, see the Australia Telescope National Facility's database: http://www.atnf.csiro.au/research/pulsar/psrcat/.}. If one includes the number
of observed X-ray binary systems \citep[e.g.,][]{liu00, liu01}, most of which 
are thought to contain a neutron star, $\lesssim2000$ NSs are known. Yet a neutron star 
is born in the Milky Way every $30$ to $100$ years, suggesting that the total 
population is $10^8 - 10^9$ objects (or roughly $1\%$ of the stars), depending on
the Galaxy's star formation history \citep[][]{neuhauser99}. Of these neutron stars, 
only the youngest will be detected as radio pulsars, provided they are aligned
favorably; after a few million years, the pulsar will have radiated away
its rotational and internal energy, and the pulses will cease \citep[][]{treves00}.
As a result, the total number of pulsars in the Milky Way is only a few 
times $10^5$; for every active pulsar there are $\sim1000$ radio-quiet 
neutron stars \citep[][]{kulkarni98}.

\citet[][]{ostriker70} proposed that some of these defunct pulsars could
reheat by accreting matter from the surrounding interstellar medium (ISM) 
through Bondi-Hoyle accretion. The ISM would need to be relatively dense, 
the NS velocity relatively low, and the NS magnetic field somewhat decayed
to allow accreting matter to reach the star's surface \citep[see also][]{treves91}.
If these conditions are met, the neutron star might emit 
approximately as a blackbody with a peak in the extreme ultraviolet/soft X-ray 
energy band \citep[][]{treves00}. For nearby NSs, this thermal emission is 
observable, and it was thought that $\ROSAT$ would detect $10^2$ 
to $10^3$ reheated isolated neutron stars (INSs), thereby potentially providing 
strong constraints on the neutron star equation of state (EOS) 
\citep[see, for example,][]{treves91, blaes93}.

Yet today the list of $\ROSAT$-detected radio-quiet isolated neutron stars contains 
just seven entries \citep[for a recent review of the so-called Magnificent Seven, see][]{haberl04}.
Recent work has suggested at least one plausible explanation for this discrepancy 
between the predicted and the observed numbers of INSs: Bondi-Hoyle accretion 
may not be an adequate mechanism for reheating large numbers of INSs, in part 
because the conditions described above are unrealistic \citep[particularly if one considers the observed pulsar velocity distribution; e.g.,][]{perna03}. 
Indeed, \citet[][]{popov00} have suggested instead that at the bright end of 
the X-ray log N-log S distribution, where most INS searches have taken place, 
the current number of known INSs is compatible with population models for young, 
cooling neutron stars. This is consistent with observations of the Magnificent Seven
suggesting they may have magnetic fields and velocities too large for Bondi-Hoyle 
accretion.

While the X-ray characteristics of the Magnificent Seven are broadly consistent 
with thermal emission, the current sample has managed to be both too small and 
too diverse in detail to definitively address the neutron star EOS. 
For example, while at least five of the Seven are X-ray pulsars \citep[][]{zane05},
the upper limit for the amplitude of X-ray pulsations in RX J1856.5$-$3754, 
the brightest known INS, is $\leq 1.3\%$ \citep[][]{burwitz03}. In addition, rather 
than bland blackbody spectra, X-ray spectroscopy of four of the Seven
has revealed unexpected broad absorption features \citep[][]{trumper05}. 
The nature and significance of each of these differences are also topics 
of current debate \citep[e.g.,][]{burwitz03, trumper05, zane05}, although 
some have argued that all the observational evidence is consistent with 
blackbody emission altered by the presence of a hydrogen atmosphere and magnetic 
fields of differing intensities \citep[][]{vankerwijk04}. Clearly, if we
are to find unifying patterns by which to disentangle the various possible
roles of magnetic fields, geometry, and atmospheres, obtaining a larger sample of
INSs is required, especially to make eventual progress towards 
understanding the fundamental questions of the neutron star EOS.

In the past, a major obstacle to finding isolated neutron stars was 
the absence of a large-area optical survey of equivalent sensitivity with which 
to identify the $>124000$ X-ray sources cataloged in the $\ROSAT$ All-Sky Survey
\citep[RASS,][]{voges99}. The availability of a suitable companion optical survey 
would allow removal of ``contaminants'' to INS searches (i.e., the bulk of more 
common X-ray-emitting subclasses: quasars, bright stars, clusters of galaxies, 
etc.), thereby narrowing the list of RASS error circles in which to search for new 
INSs. \citet{rutledge03} attempted to identify candidate INSs from among the $19000$
RASS sources in the Bright Source Catalog \citep[BSC,][]{voges99} by eliminating 
matches to the United States Naval Observatory (USNO)
A$2.0$ optical catalog, but in the interim the Sloan Digital 
Sky Survey \citep[SDSS;][]{york00} has emerged as a more powerful companion optical 
survey to RASS, especially for extending the search for new INSs to the RASS 
Faint Source Catalog \citep[FSC,][]{fsc}.

Here we describe a program to identify the best candidate isolated neutron stars from
correlations of the RASS Bright and Faint Source Catalogs and an early version
of the SDSS Data Release 4 \citep[DR4;][]{DR4paper}. In the following 
section we outline the properties of the two surveys. Section 3 describes the 
method used to select our candidate fields, and in section 4 we describe the 
properties of the individual candidate fields identified by this program. Section 5 
is a discussion of our results and includes a comparison of our method and that 
of \citet[][]{rutledge03}, an earlier search for INSs that also used the RASS.\footnote{\citet[][]{chiere05} have recently published four $\ROSAT$-detected X-ray sources without optical counterparts to the Guide Star Catalog faint limit of $19 - 23$ mag (depending on the optical band and the source position on the sky). These are High Resolution Imager observations, and the area of sky covered and the X-ray error circles are both much smaller than for the RASS.}
We conclude in section 6.

\section{          RASS and SDSS: A Match Made in the Heavens         }
The $\ROSAT$ All-Sky Survey (RASS) was the first of its kind in soft X rays 
($\sim0.1 - 2.4$~keV).
Using the Position Sensitive Proportional Counter, $\ROSAT$ imaged the sky with
exposures of lengths ranging from $\sim400$ to $\sim40000$ s at the ecliptic 
equator and poles, respectively, with $99.7\%$ of the sky observed in exposures 
at least $50$ seconds long \citep[][]{voges99}. The typical limiting sensitivity 
of the resulting RASS catalog is a few times $10^{-13}$~ergs~cm$^{-2}$~s$^{-1}$, 
and more than $124000$ sources are included when one merges the RASS Bright 
and Faint Source Catalogs \citep[][]{voges99,fsc}.

The Sloan Digital Sky Survey provides a uniform optical 
photometric and spectroscopic dataset with which to correlate the RASS catalog. 
SDSS is currently mapping the sky at optical wavelengths using 
a dedicated 2.5~m telescope at the Apache Point Observatory, New Mexico, and 
producing homogeneous five color {\it u,g,r,i,z} CCD images to a depth of $r\sim22.5$ 
\citep[][]{fukugita, gunn, hogg01, smith02, gunn05}, with associated photometry accurate to 
$0.02$ magnitudes \citep[][]{zeljko04}. Astrometric accuracy is better than $0.1\asec$ 
per coordinate (rms) for sources with $r<20.5$ \citep[][]{pier}; morphological
information drawn from SDSS images allows for reliable star/galaxy separation to 
$r\sim21.5$ \citep[][]{lupton02}. The survey's coverage of $\sim10^4$~deg$^2$ 
around the north Galactic cap and of $\sim200$~deg$^2$ in the southern Galactic 
hemisphere will result in photometric measurements for over $10^8$ stars and a 
similar number of galaxies. SDSS will also obtain spectra for 
$10^6$ galaxies and $10^5$ quasars. The fourth public Data 
Release (DR4) includes photometric data for $6670$~deg$^2$ of sky, and catalogs 
$1.8 \times 10^8$ objects \citep[][]{DR4paper}.

Largely by coincidence, the RASS and SDSS are extremely well-matched, making
SDSS an ideal tool for identifying large numbers of $\ROSAT$ sources
\citep[e.g.,][]{anderson03}. In particular, if one considers the known range 
of $f_x/f_{opt}$ for common X-ray emitters, even the faintest optical 
counterparts to typical RASS sources are bright enough to be detected in 
the SDSS photometric survey and targeted for SDSS spectroscopy.
For the typical classes of X-ray emitters, including normal stars, normal
galaxies, quasars, and BL Lacs, the highest X-ray-to-optical flux ratios
have log $(f_x/f_{opt})$ values of about $-1$, $0$, $+1$, and $+1.5$, 
respectively \citep[e.g.,][]{stocke91,zickgraf03}. Given the RASS flux limit 
quoted above, this implies that a faint optical counterpart in each of 
these categories of typical X-ray counterparts will have $m \lesssim 15, 17, 20$, 
and $21$, respectively, and therefore that SDSS will obtain accurate photometry 
for the vast majority of RASS counterparts within its footprint.
(We use the \citet{macca88} formula for calculations of 
log~$(f_x/f_{opt})$, and substitute $g$ magnitudes for $m_V$\footnote{While $g$ and $m_V$ are not equal, the color-dependent difference between the two is relatively small \citep[$g = m_V + 0.05$ for a typical low-redshift quasar with (B$-$V) $= 0.3$;][]{fukugita}.}). Furthermore, 
at these magnitudes the SDSS spectroscopic survey will frequently obtain good 
signal-to-noise spectra for targeted suspected counterparts, allowing for 
confident identifications.

\section{ Using SDSS to Identify Optically Blank RASS Fields }

Isolated neutron stars, however, are not among the typical classes of X-ray 
emitters, and have anomalously large log $(f_x/f_{opt})$ values compared to other
X-ray sources due to their optical faintness. Of the Magnificent Seven, four 
have suggested optical counterparts with
$m_V$ between $25.8$ and $28.7$ and associated log~$(f_x/f_V)$ values between
$4.4$ and $5.0$ \citep[][]{kaplan03}. Clearly, an optical counterpart to an INS is
unlikely to be found using SDSS. Rather, we use SDSS to search 
for RASS fields devoid of plausible optical counterparts to the SDSS $m\sim22$ 
limit. Such error circles host X-ray sources with such extreme $f_x / f_{opt}$ 
ratios that an INS becomes a plausible identification.

We select the RASS BSC and FSC objects within the SDSS DR4 footprint 
by querying the DR4 database for a complete list of SDSS field positions. 
SDSS fields are $2048 \times 1489$ pixels and consist of 
the frames in the five SDSS filters for the same part of the sky. An SDSS field 
is in some sense the survey's smallest imaging unit, in that all elements 
of a given field are processed by the photometric pipeline at one time. 
Matching the RASS positions with those of the $\sim250000$ DR4 fields 
obtained, we find that $\sim22700$ RASS sources are within the area defined 
by the DR4 fields, which covers $6670$~deg$^2$. Since there are roughly three 
RASS sources per deg$^2$, this number of X-ray sources in the DR4 footprint 
is consistent with the number of RASS sources expected from a simple surface 
density argument. Unlike other INS candidate searches \citep[e.g,][]{rutledge03},
we do not apply a cut based on the measured X-ray hardness 
ratios \citep[HR1 and HR2; see][]{voges99}. 
The Magnificent Seven have HR1 ratios that range from $-1$ to $0$, not a strong
constraint when the possible range is $-1$ to $1$. While the HR2 ratio may provide
a better tool with which to identify soft X-ray emitters, the uncertainties associated 
with the count rates for RASS faint X-ray sources make this ratio
practically undetermined for many of the sources we consider here. 

To find the best isolated neutron star candidate fields, we search the 
$22700$ RASS sources for those with small X-ray positional 
uncertainties and select the $\sim9500$ with quoted positional errors ($1\sigma$) 
smaller than $15\asec$ (the median RASS positional error for this 
sample is $13\asec$). In identifying counterparts to these sources in SDSS and 
other catalogs (and thereby eliminating them), we generally restrict our search 
to objects within a disk centered on the RASS position 
and of radius either $1\amin$ or $4$ times the quoted X-ray positional error.
\citet[][]{voges99} provide one empirical distribution of the positional 
offsets of optical counterparts relative to the quoted BSC positional 
errors \citep[see their Fig.~8, compiled from correlations with the 
TYCHO catalog;][]{hog}. An examination of their most reliable matches indicates 
that this distribution is not adequately described overall by a  
two-dimensional Gaussian, and that a one-dimensional Gaussian may be 
a better fit at larger multiples of the quoted positional error. We 
therefore estimate that among the $9500$ sources we consider further, fewer 
than 1 is expected to have a counterpart with a positional offset larger than 4
times the $\ROSAT$ X-ray positional error. For simplicity, we describe 
this search radius in the rest of the text as equal to $4$ $p.e.$ 
(for positional error), and the associated error circle as the $4$ $p.e.$ error
circle.

Previous work suggests that roughly one third of the $\sim9500$ sources with small
positional errors are quasars and that another third are bright stars 
\citep[e.g.,][]{zickgraf03}. We therefore match the $9500$ sources with small 
positional errors to the most recent SDSS catalog of $>4000$ spectroscopically 
identified RASS quasars \citep[e.g.,][]{anderson03}. 
We take $1\amin$ as our matching radius, meaning that any X-ray source with 
a spectroscopically confirmed quasar within an error circle of radius 
$\geq4$ $p.e.$ is eliminated from further consideration.
Similarly, we use the SDSS DR4 photometric 
catalog to eliminate RASS fields with a $g < 15$ mag object\footnote{In querying the database for photometry, we request both PSF and model magnitudes, and make our cuts based on both. Typically, PSF fitting provides better estimates of isolated star magnitudes, while model fitting is best for galaxies. See \citet[][]{stoughton02}.} within $1\amin$.
When querying the DR4 database, we request ``primary'' photometry, which requires
that objects have a single entry in the database, that they not be deblended, and 
that they fall within the survey boundaries \citep[for details, see][]{stoughton02}.
The median count rate for our sample of RASS sources with small positional errors
is $0.034$ counts s$^{-1}$, so that $g < 15$ objects have 
log~$(f_x/f_g) \lesssim~-1.1$ and are most probably the X-ray source 
counterparts \citep[see Table~1 in][]{stocke91}. Both of these cuts
are extremely conservative, as they are applied without specific $f_x/f_{opt}$
restrictions and extend to a large positional offset for
each source. Still, a confirmed quasar, or a bright star or galaxy, even 
with an atypical $f_x/f_{opt}$ and at a large positional offset, might be a more
plausible identification than an INS. Roughly half of the $9500$ originally 
selected sources with small positional errors remain at the end of this stage 
of our algorithm.

To reduce the number further, we use the DR4 database to eliminate X-ray error 
circles with any UV-excess objects. These are objects satisfying $u-g < 0.6$ 
and $u \leq 22.0$, the SDSS $95\%$ completeness limit \citep[][]{stoughton02}; 
for the most part, these are candidate (photometric) quasars 
\citep[e.g.,][]{richards02}, but this cut also identifies and removes white dwarfs and 
cataclysmic variables/X-ray binaries. Here we calculate the separation between 
each such object and the RASS source, and eliminate those fields where a UV-excess 
object falls within the $4$ $p.e.$ 
error circle of the associated X-ray source. 
Finally, because the occasional pathologically bland quasar 
\citep[a few percent of all cases;][]{vandenberk01} or a quasar in a 
selected redshift range can have colors consistent with those of normal stars 
and cannot easily be identified using SDSS-color cuts, we remove fields 
with objects that have quasar-like X-ray-to-optical flux ratios \citep[defined as 
log~$(f_x/f_g) \leq 1.2$, the typical upper limit for AGN given 
by][]{stocke91} {\it regardless of their optical colors}. 
We eliminate all fields where such an object with $g \leq 22.2$ 
(the $95\%$ completeness limit in that band) is cataloged within $4$ $p.e.$ 
of the X-ray position. This removes about four-fifths of the remaining sources,
leaving us with $410$ X-ray sources, about $4\%$ of our initial sample of 
RASS sources with small positional errors\footnote{This cut is sufficiently stringent that it may render some of the previous steps unnecessary. However, it would not alone eliminate spectroscopically confirmed SDSS BL Lacs, for example, a significant fraction of which do have log~$(f_x/f_g) > 1.2$ \citep[][]{anderson03}.}.

We then require, when matching these sources to the SDSS catalog, that SDSS primary 
photometry within the $4$ $p.e.$ error circle be available, thereby eliminating 
false positives--fields that would otherwise be defined as optically blank at 
this stage of the algorithm only because there is no reliable SDSS 
photometry for them. Over $80\%$ of the $410$ fields lack SDSS photometry; 
this is frequently because of the presence of a saturating star or because 
the RASS source falls on the edge of the DR4 footprint. The remaining $74$ 
RASS sources are then fed to SIMBAD and NED, with which we eliminate $19$. 
Roughly half of these sources are cataloged clusters within $1\amin$, while the 
rest include known BL~Lacs or bright 2MASS galaxies within the $4$ $p.e.$ error
circle. We also eliminate the $22$ X-ray error circles 
with a cataloged National Radio Astronomy Observatory (NRAO) Very Large Array (VLA) Sky 
Survey \citep[NVSS;][]{nvss} or VLA Faint Images of the Radio Sky at 
Twenty-Centimeters \citep[FIRST;][]{first} source within $4$ $p.e.$ of the 
associated X-ray position; this is intended especially to eliminate uncataloged
BL~Lac candidates. These steps reduce the list of candidates to $33$ RASS X-ray 
error circles.

Visual inspection of the SDSS images of these remaining $33$ fields finds three cases 
with an obvious candidate optical counterpart to the X-ray source that has somehow 
evaded our algorithm. In two cases the most likely explanation is the absence 
of primary SDSS photometry for a bright star present in the center of one field 
and for a galaxy in the center of the other. In the third case SDSS 
spectroscopy of a $g = 20.44$ object reveals it to be an emission line galaxy 
and therefore a conceivable (though unusual) X-ray source. 
We eliminate these fields and are left with a list of $30$ X-ray error circles 
that are optically blank (devoid of plausible optical counterparts)
at the SDSS catalog level as defined by our algorithm.

Visual inspection of the {\it ROSAT} hard and soft band images of these $30$ 
fields allows us to eliminate nine RASS sources from further consideration. 
These eliminated sources include possible artifacts and extended or very uncertain 
X-ray detections. Among the RASS sources eliminated at this stage is 
1RXS J115309.7$+$545636, one of the sources observed with the {\it Chandra 
X-ray Observatory} as a candidate INS by \citet{rutledge03}. Their 
observations confirmed that 1RXS J115309.7$+$545636 is in fact not an INS (see \S 5.1).

Three additional RASS sources were set aside at this stage. Their 
quoted positional error of $6\asec$ appears to be an underestimate if one
considers the count rates and exposure lengths for these sources 
(see Appendix for further discussion).

Some fraction of the remaining X-ray sources are likely to be associated with
optically faint clusters of galaxies. We therefore correlate our remaining X-ray
positions with a catalog of optically selected SDSS clusters (J.~Annis 2005, private 
communication). Two of our X-ray sources fall outside of the cluster catalog's 
footprint, and were therefore not subject to this cluster analysis; see Table~1. 
The optical clusters are described in part by their $n_{gal}$, the number of 
red sequence galaxies brighter than $0.5 L_\star$ within a 1~Mpc radius volume, 
with SDSS colors providing a photometric estimate of the redshifts.
If there is a candidate cluster with $n_{gal} \geq 3$ with an offset $\leq 1\amin$
from the X-ray position, we eliminate the X-ray source as a candidate INS. 
This removes two RASS sources from consideration and is consistent with our 
elimination, earlier in our algorithm, of cases with a cataloged NED/SIMBAD 
cluster within $1\amin$.

In identifying clusters at larger angular separations as the likely RASS X-ray 
source, we take $5\amin$ as our maximum separation for considering 
matches. There is a drop off in the surface density of candidate SDSS optical clusters
at that separation. We choose $n_{gal} \geq 6$ as our richness criterion 
for considering a cluster to be the likely X-ray source, in these high positional 
offset cases, and thereby eliminate another five RASS error circles as 
candidate clusters (see Appendix for further discussion of these candidate clusters).

These steps winnowed the original list of $9500$ RASS sources in the SDSS 
DR4 footprint to $11$ X-ray sources that are bereft of plausible 
counterparts and are therefore candidate blank field X-ray sources.
It is highly reassuring that among these surviving blank field 
X-ray sources is the field containing RX J$1605.3+3249$, the only previously 
known isolated neutron star in the SDSS DR4 footprint. We argue 
in \S 5.1 that our selection using SDSS is an order of magnitude more stringent 
than the \citet{rutledge03} hallmark blank field search for INSs.

\section{       Properties of the Candidate Isolated Neutron Star Fields }
The $11$ fields discussed in this section are those which survived our winnowing 
algorithm and are most likely to harbor either isolated neutron stars or some 
other rare and exotic X-ray emitter such as radio-quiet BL Lacs, obscured AGN, 
dark clusters, etc. They include one previously rejected candidate
INS field \citep{rutledge03}, as well as the field of RX J$1605.3+3249$, the only
known INS falling within the SDSS DR4 footprint. Fig.~\ref{fields1} is 
a mosaic of the SDSS composite $g,r,i$ images of all $11$ of the candidate fields
with the RASS $4$ $p.e.$ error circles superimposed. 
Table~1 includes the {\ROSAT} parameters for all of these RASS sources 
\citep[counts s$^{-1}$, detection likelihood, and exposure time; see][]{voges99}, as well as the $g$ magnitude of the {\it brightest} SDSS object 
within the $4$ $p.e.$ RASS error circle and the corresponding {\it minimum} 
log~$(f_x/f_g)$ for the optical counterpart to the X-ray source.

Below we provide additional information about several of these X-ray sources:  
those that might have viable optical counterparts at unexpectedly large positional 
offsets, and those previously identified in the literature as INS candidates.

\begin{table}[ht]
\small
\begin{center}
\begin{tabular}{lcccc|cc}
\hline
\multicolumn{5}{c|}{{\it ROSAT}} & \multicolumn{2}{c}{SDSS}\\
\multicolumn{1}{c}{Source name} & & Count rate & Detection & Exp. & $g$ & Min. \\
\multicolumn{1}{c}{1RXS J} & $1\sigma$ & $10^{-2}$ cps & likelihood & s & mag & log $(f_x/f_g)$ \\
\hline
\hline
003413.7$-$010134\tablenotemark{a} & $14\asec$ & $1.3\pm0.6$ & 10 & 630 & $23.22$ & $1.8$ \\
013630.4$+$004226\tablenotemark{a} & $13\asec$ & $2.5\pm1.0$ & 16 & 283 & $21.82$ & $1.5$ \\
092310.1$+$275448 & $14\asec$ & $2.5\pm1.0$ & 14 & 407 & $21.74$ & $1.5$ \\
103415.1$+$435402 & $14\asec$ & $1.7\pm0.8$ & 9  & 502 & $21.66$ & $1.3$ \\
110219.6$+$022836 & $15\asec$ & $1.8\pm0.8$ & 11 & 423 & $21.80$ & $1.3$ \\
122344.6$+$373015 & $15\asec$ & $2.8\pm1.1$ & 11 & 490 & $21.97$ & $1.6$ \\
130547.2$+$641252\tablenotemark{b} & $9\asec$  & $16.7\pm2.1$ & 122 & 544 & $22.05$ & $2.4$ \\
131400.1$+$072312 & $15\asec$ & $1.8\pm1.0$ & 8  & 333 & $21.69$ & $1.3$ \\
141428.5$+$601707 & $14\asec$ & $1.4\pm0.7$ & 9  & 630 & $21.99$ & $1.3$ \\
151855.1$+$355543 & $9\asec$  & $3.3\pm0.9$ & 32 & 518 & $21.04$ & $1.3$ \\
160518.8$+$324907\tablenotemark{c} & $7\asec$  & $87.5\pm4.1$ & 1140 & 566 & $22.80$ & $3.4$ \\
\hline
\end{tabular}
\tablenotetext{a}{Outside of optical cluster catalog footprint}
\tablenotetext{b}{Rejected as an INS by \citet{rutledge03}}
\tablenotetext{c}{Known INS}
\caption{X-ray and optical data for the $11$ candidate isolated neutron star fields.}
\end{center}
\end{table}

\subsection{New candidate INS fields with other possible counterparts at large offsets}
There are two candidate INS fields in which a known quasar or a bright star
lies just outside the $4$ $p.e.$ RASS error circle. Although we expect 
$< 1$ case of an optical counterpart being found at such large positional offsets
from our entire starting set of $9500$ RASS X-ray sources, in this section we call 
special cautionary attention to these cases. Good angular resolution X-ray
images would quickly resolve such issues definitively. These two cases are:

\begin{figure}[th] 
\centerline{\includegraphics[width=.99\columnwidth]{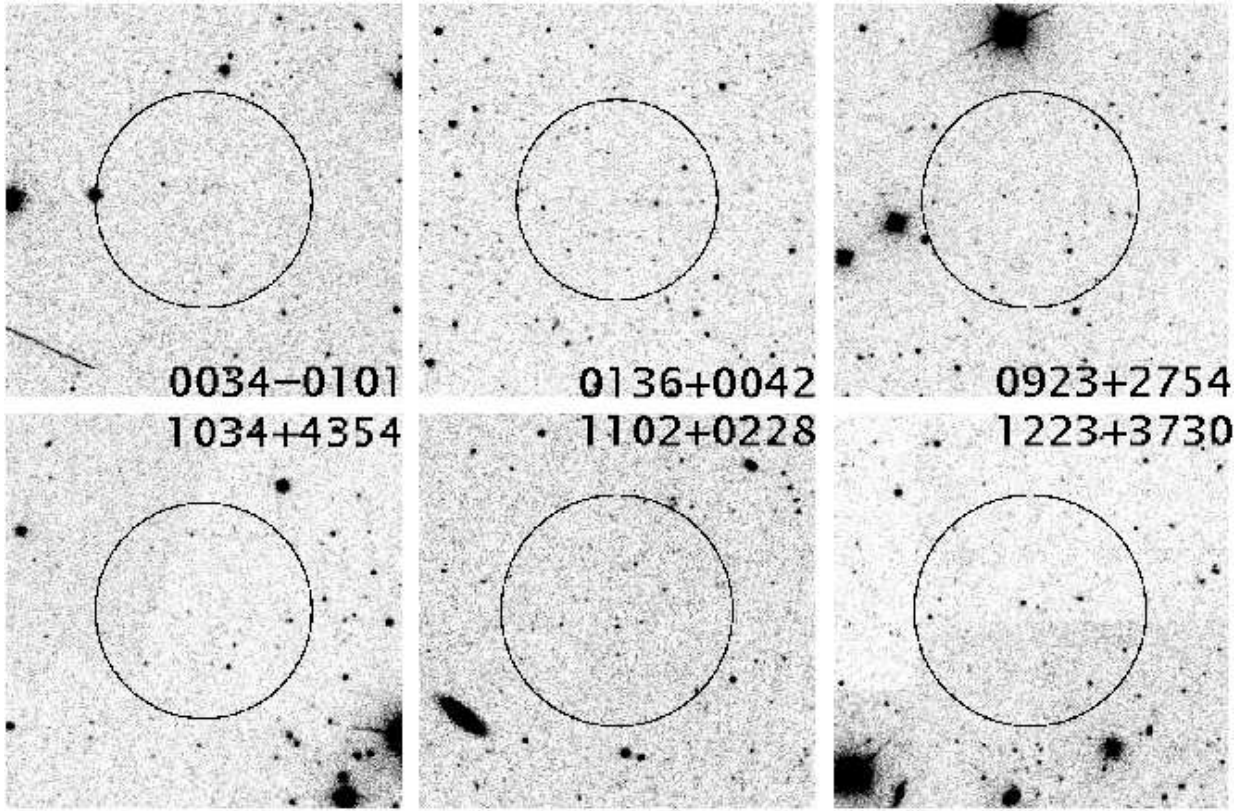}} 
\caption{SDSS composite $g,r,i$ images of our best isolated neutron star 
candidate fields, with the stretch being the same for all. The $4$ $p.e.$
RASS positional error circle is shown in each image. The brightest SDSS object 
seen within any of the error circles is $g = 21.66$ mag. North is up and East 
is to the left and the images are roughly $3.5\amin$ on a side.}
\label{fields1} \end{figure}

\begin{itemize}
\item {\bf 1RXS J003413.7$-$010134} The $4$ $p.e.$ error circle just barely excludes
a $g = 16.75$ star for which we obtained a spectrum with the $3.5$-m telescope 
at Apache Point Observatory (APO), New Mexico. This spectrum is of that of a 
G star with no emission; the star's 
log~$(f_x/f_g)$ of $-0.8$ is unlikely for G stars, whose (log) flux ratios are
typically between $-4.3$ and $-2.4$ \citep[see Table~1 of][]{stocke91}, suggesting
that it is probably not the X-ray source.

In addition, we note the presence of a spectroscopically confirmed quasar,
SDSS J003413.04$-$010026.8, $1.13\amin$ ($4.8 \times$ the quoted RASS positional 
error) from this source. This $g = 17.20$ quasar has a log~$(f_x/f_g) = -0.63$, 
within the range for AGN given by \citet{stocke91} of $-1$ to $1.2$. 

\begin{figure}[th]
\figurenum{1}
\centerline{\includegraphics[width=.99\columnwidth]{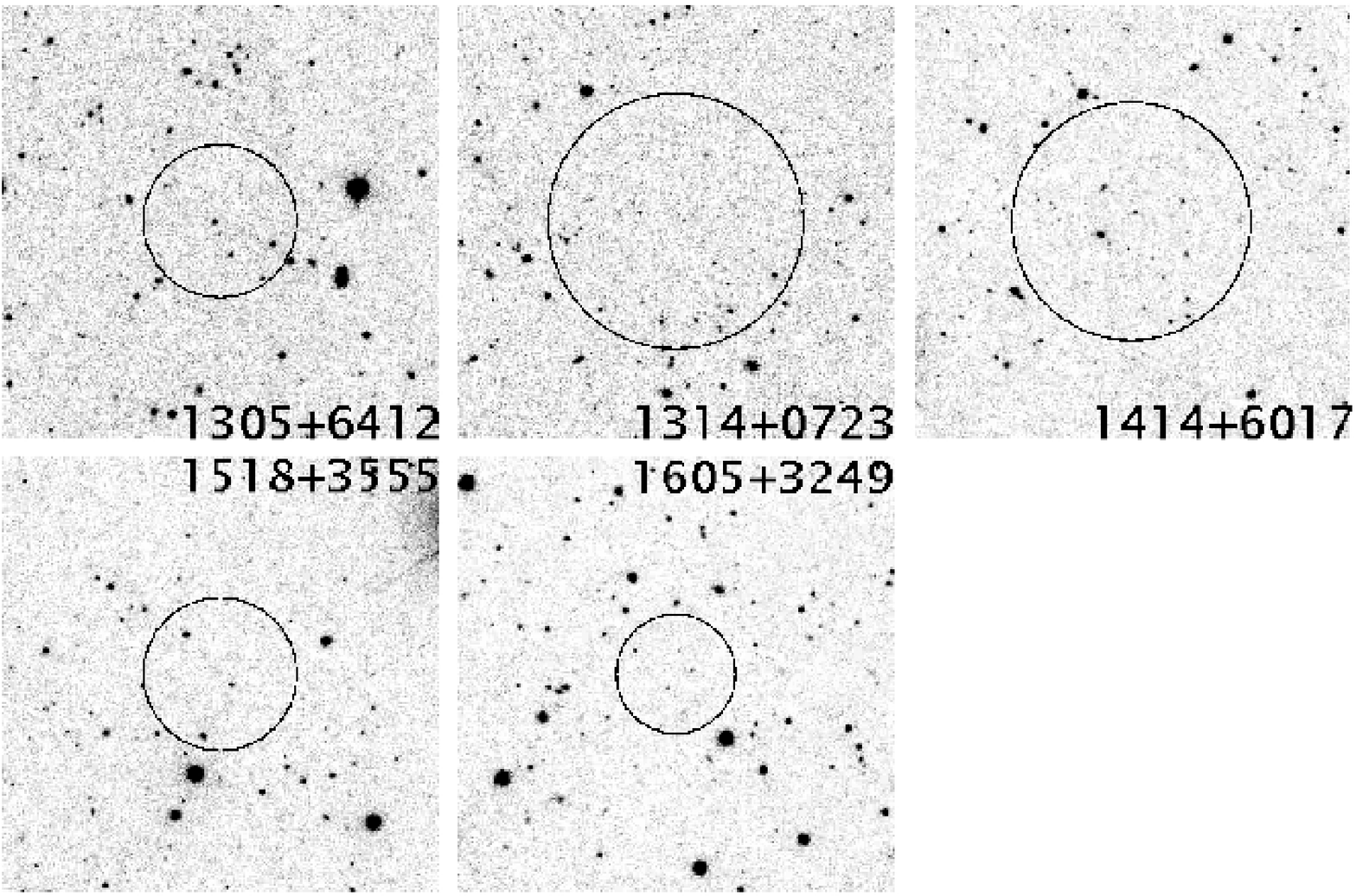}} 
\caption{{\it (Continued)} J130547.2$+$641252 was rejected as an INS by \citet[][]{rutledge03}. J160518.8$+$324907 is a known INS. The brightest SDSS object within any of the $4$ $p.e.$ error circles is $g = 21.04$ mag.}
\end{figure}

\item {\bf 1RXS J141428.5$+$601707} A spectroscopically confirmed quasar, 
SDSS J141431.67$+$601807.2, lies $1.09\amin$ ($4.7 \times$ the quoted RASS positional
error) from this source. This quasar has $g = 17.82$ mag, so that its 
log~$(f_x/f_g) = -0.36$, within the range for AGN given by \citet{stocke91}.
\end{itemize}

\subsection{Previously known candidate INS fields}

There are two previously suggested candidate isolated neutron stars that are also 
included among our $11$ candidates. One is a confirmed INS, while the other
was later refuted as an INS:

\begin{itemize}
\item {\bf 1RXS J130547.2$+$641252} This source was proposed and rejected 
as a candidate INS by \citet{rutledge03} because of its X-ray variability 
(see \S5.1.1 for further discussion).

\item {\bf 1RXS J160518.8$+$324907} This is the only previously known INS 
in the SDSS DR4 footprint. It is very reassuring that this confirmed case is
recovered by our algorithm.
\end{itemize}

\section{                  Discussion          }

\subsection{Comparison with Previous Work}
An important effort involving RASS/optical selection of candidate isolated 
neutron stars was that of \citet{rutledge03}. \citet{rutledge03} identified 
candidate INS fields by correlating the RASS Bright Source Catalog 
with NVSS, the {\it Infrared Astronomical Satellite} Point Source Catalog, and 
the United States Naval Observatory A2.0 optical catalog. They obtained 
a list of $32$ candidate blank field RASS sources (including two known INSs), from which 
they selected eight for {\it Chandra X-ray Observatory} observations. None
of these eight sources was found to be an INS \citep[][]{rutledge03}.

The current availability of both SDSS spectroscopy and much deeper SDSS
photometric data--two or three magnitudes fainter than USNO A$2.0$--permits
us to invoke a much more stringent set of selection criteria. To compare our
method and that of Rutledge et al., we discuss the properties of the $11$ 
candidate fields from their original list of $32$ that fall within the SDSS 
DR4 footprint, as well as those of the six of their eight {\it Chandra} targets that 
do not fall within the SDSS footprint, but for which \citet{rutledge03} provide 
Digitized Sky Survey (DSS) data. Finally, we speculate on how the remaining $15$ 
Rutledge et al.\ blank field candidates would fare at the hands of our 
algorithm {\it if} they were in the SDSS DR4 footprint.

\subsubsection{\cite{rutledge03} sources within SDSS DR4}
Of the initial $32$ Rutledge et al.\ candidate sources, $11$ fall within the 
DR4 footprint, including the previously known INS, RX J$1605.3+3249$. 
All $11$ candidates were processed by our winnowing algorithm in a double-blind,
end-to-end fashion. Reassuringly, RX J$1605.3+3249$ survives our algorithmic 
selection, and is among our $11$ INS candidates.

Six of the remaining $10$ \citet{rutledge03} sources, for which they 
identified ``ordinary'' optical counterparts, were eliminated early on 
by our algorithm. One 
(1RXS J091010.2$+$481317) disappears from our list of candidate sources when 
matched against spectroscopically confirmed SDSS quasars, three
(1RXS J104710.3$+$633522, J130402.8$+$353316, J130753.6$+$535137) are eliminated 
because of the presence of candidate photometric SDSS quasars\footnote{This eliminates the fields with known cataclysmic variables (1RXS J104710.3$+$633522 and J130753.6$+$535137), as these also have $u - g \leq 0.6$.}, and 
two (1RXS J123319.0$+$090110, J130034.2$+$054111) fail our test for sources with 
no objects brighter than $g = 15$ within $1\amin$. A seventh source,
1RXS J094432.8$+$573544 is eliminated when our X-ray sources are matched to 
the radio catalogs; it also has an SDSS spectrum, which suggests that it is a BL Lac.
 
Two of the three other sources survive to the last stages of our algorithm before being
eliminated. Visual inspection of the RASS images of 1RXS 115309.7$+$545636 reveals 
it to be an extended source, and it is therefore eliminated as a candidate INS by
our algorithm. 1RXS J145234.9$+$323536 is identified as a candidate optical cluster and
also removed from our list of INS candidates (see Appendix). \citet{rutledge03} 
obtained {\it Chandra} observations of both of these sources and did not detect 
an X-ray source in either case.

Only one of the remaining $10$ Rutledge et al.\ sources, 
1RXS J130547.2$+$641252, survives to 
make our list of the best candidate isolated neutron stars. \citet{rutledge03} rejected
this source as a candidate based on its X-ray variability, measured by comparing its 
RASS data to observations of the same source with the High Resolution Imager on $\ROSAT$.
We queried the {\it Chandra} and {\it XMM-Newton X-ray Observatory} lists of 
observed targets as well as the various $\ROSAT$ catalogs and found that 
none of our other INS candidates 
has the complementary X-ray observations required for the detection of such 
variable or transient sources. We therefore cannot discount the  
possible contamination of our candidate list by such sources.
 
In summary, of the $11$ Rutledge et al.\ candidate isolated neutron star fields that 
fall within the DR4 footprint, only two survive our winnowing process. One is a likely
transient or variable source, previously discounted as an INS via {\it Chandra} 
observations by \citet{rutledge03}. The other is a successful recovery of the 
one previously confirmed INS in the SDSS DR4 imaging area, RX J$1605.3+3249$. 
Our algorithm is therefore significantly more efficient at removing contaminants 
from our candidate list while simultaneously recovering the only previously 
confirmed INS in the SDSS DR4 footprint.

\subsubsection{Sources for which \citet{rutledge03} provide DSS data}
\citet{rutledge03} also obtained {\it Chandra} observations for six sources 
that do not fall within the SDSS DR4 footprint. However, Rutledge et al.\ do 
discuss the DSS photometric properties of these six sources: 
for three they find fairly bright likely optical counterparts, and for 
three they find counterparts near the DSS faint limit, 
all offset from the RASS source by less that $3 \times$ the quoted 
positional error. While these six sources are not in the DR4 footprint, 
it is almost certain that had they been, none would have made our list 
of candidate isolated neutron stars. 

In three of the X-ray error circles (1RXS J024528.9$+$262039, J132833.1$-$365425, 
and J163910.7$+$565637), \citet{rutledge03} identify the likely optical counterparts 
as two B~$\sim15$ late type stars and a $B = 17.8$, $z = 1.65$ quasar, respectively. 
With SDSS data, the magnitudes of the late type stars and/or their 
$f_x/f_{opt}$ ratios \citep[$\sim-0.5$ and $\sim-0.9$, respectively, within the range
for M stars in][]{stocke91} would very likely have caused our algorithm to eliminate
these error circles, and perhaps even suggested the proper identifications. 
The low redshift quasar's colors might have been unusual, causing our algorithm to
remove it from our candidate list early on; if not, its $f_x/f_{opt}$ ratio ($\sim0.2$)
likely would also have caused it to be eliminated by our algorithm (an SDSS 
spectrum might also have been available).

In the other three cases (1RXS J020317.5$-$243832, 145010.6$+$655944, and 
122940.6$+$181645), SDSS photometry would be available, as the SDSS faintness limit 
is $\sim2-3$ magnitudes deeper than the DSS limit, depending on the band. 
Rutledge et al.\ suggest a faint cataclysmic variable (CV) 
and two faint AGNs as the optical
counterparts to these X-ray sources. The colors of the objects, again along 
with their $f_x/f_{opt}$ ratios, would probably have disqualified these 
fields from further consideration.

Our algorithm therefore would probably have eliminated all six of these
additional sources observed with {\it Chandra} by \citet{rutledge03},
none of which were confirmed as an INS.

\subsubsection{Other \cite{rutledge03} sources}
There remain $15$ sources that \citet{rutledge03} initially considered as candidate 
isolated neutron stars, but ultimately rejected in post-algorithmic screening. 
The optical content of these RASS error circles as described by Rutledge et al., 
or as directly determined from DSS images by us, is such as to virtually guarantee 
that $14$ would be rejected by our algorithm {\it if} they fell within the 
SDSS footprint. These fields include $12$ with bright stars, a CV in the 
globular cluster M3, and a known Seyfert 1 galaxy. The $15^{\rm th}$ source is 
another known INS, RX J1308.6$+$2127. 

In summary, these comparisons verify that our algorithm, relying especially on
the greater optical photometric depth of SDSS, successfully recovers the only 
previously known INS in our survey area, and also rejects an order of magnitude more
contaminating RASS error circles than the \citet{rutledge03} 
search for isolated neutron stars.

\subsection{Comparison With Galactic INS Population Models}
The current dearth of candidate isolated neutron stars beyond the Magnificent 
Seven has led to several efforts to rethink the expected population of INSs
within the Galaxy. In particular, \citet{popov00} compared the space density of 
accreting (i.e., reheated and old) isolated neutron stars to that of cooling (i.e., 
young) neutron stars, using a number of assumptions about the Galactic neutron star
birth rate, the large-scale distribution of gas in the interstellar medium, the
cooling time for a newborn neutron star, etc.

\citet{popov00} find that at the bright end ($\ROSAT$ count 
rates $\geq 0.1$ counts s$^{-1}$, L$_x$ $\sim 10^{29} - 10^{30}$ 
ergs~s$^{-1}$), the predicted population of INSs is essentially just the small 
number of young neutron stars seen at an early enough evolutionary stage--the 
first $10^6$ years of their lives--to still be quite hot. They also
find that these coolers are typically three orders of magnitude brighter than 
accreting, older isolated neutron stars. However, the total number of predicted
accretors is about two orders of magnitude larger than that of coolers, 
so that at lower count rates/X-ray fluxes, the number of accretors
is comparable to that of coolers. Indeed, \citet{popov00} predict that at fluxes
below $\sim10^{-13}$~ergs~cm$^{-2}$~s$^{-1}$, the number of accretors exceeds 
the number of coolers. The overall \citet{popov00} prediction for the Galactic 
INS population is consistent with current observations, if one assumes that 
the Magnificent Seven are indeed all young, bright coolers. Interestingly, 
it also suggests that at the fainter flux limits of the RASS Faint Source 
Catalog, there are a significant number of as-of-yet undetected isolated 
neutron stars.

Of the $\sim9500$ RASS X-ray sources with small positional errors 
from which we selected our candidate INSs, $80\%$ have count rates 
$\geq 0.017$ counts s$^{-1}$, and we adopt this value as a rough lower limit to 
our search sensitivity. This corresponds to the peak in the cumulative
distribution of count rates for our sample, and is also consistent with 
the value of the flux 
for which \citet[][in preparation]{shen} quote a $50\%$ completeness level 
for RASS detections in the SDSS Data Release 1 area. \citet{popov00} predict 
that the total number of isolated neutron stars with count rates greater than
this limit should be $3 - 10$ steradian$^{-1}$, or $40 - 125$ over the 
entire sky. Based on their models, a naive prediction is that the SDSS DR4 
area contains $5 - 20$ isolated neutron stars, a range consistent 
with the number of new candidates we identify here. We note, however, that
our list is not complete: it is very likely that good INS candidates 
were lost because of their chance proximity to unrelated SDSS objects that 
caused our algorithm to eliminate those RASS error circles from 
consideration.

\section{                   Conclusion                    }

In an effort to expand the sample of known isolated neutron stars, we have 
developed a selection algorithm based on a cross-correlation of the RASS and SDSS 
data to identify X-ray error circles devoid of plausible optical counterparts. 
We use SDSS spectroscopy and, especially, deep SDSS DR4 photometric data 
to quantitatively characterize the $11$ RASS fields that survive our 
winnowing algorithm as optically blank to the SDSS $g \sim 22$~mag
faint limit. Our search is an order of magnitude more selective than similar previous 
searches for optically blank RASS error circles; in selecting our INS candidates,
we have excluded $99.9\%$ of the RASS error circles in our initial sample.

The $11$ RASS fields we identify as potentially hosting an INS include the only 
confirmed INS in the DR4 footprint, RX J$1605.3+3249$, along with 
1RXS J130547.2$+$641252, previously considered as an INS candidate and
rejected on the basis of {\it Chandra} observations \citep[][]{rutledge03}.
The remaining nine new candidates may host INSs or other 
similarly exotic X-ray sources, such as unusual X-ray binaries, high-redshift
quasars, dark clusters of galaxies, type 2 quasars, or extreme BL~Lacs 
\citep[e.g.,][]{chiere05}. 

We note that the number of candidates we find is consistent with the 
predictions from recent INS population models for the number expected in the 
SDSS DR4 footprint. Planned {\it Chandra} follow-up observations of these 
candidate fields will help confirm whether they contain isolated neutron 
stars or some alternate exotic X-ray emitters. At the minimum, our sample 
may help increase the diversity of neutron stars available for study.

\begin{acknowledgements}
We thank Kevin Covey, Anil Seth, and the observing specialists at the telescope
for their assistance with the APO observations. We thank the referee, Marten van 
Kerkwijk, for suggestions that improved the manuscript.  

Funding for the creation and distribution of the SDSS Archive has
been provided by the Alfred P. Sloan Foundation, the Participating
Institutions, the National Aeronautics and Space Administration, the
National Science Foundation, the U.S. Department of Energy, the
Japanese Monbukagakusho, and the Max Planck Society. The SDSS Web
site is http://www.sdss.org/.

The SDSS is managed by the Astrophysical Research Consortium (ARC) for
the Participating Institutions. The Participating Institutions are
the University of Chicago, Fermilab, the Institute for Advanced Study,
the Japan Participation Group, The Johns Hopkins University, the
Korean Scientist Group, Los Alamos National Laboratory, the
Max-Planck-Institute for Astronomy (MPIA), the Max-Planck-Institute
for Astrophysics (MPA), New Mexico State University, the University of
Pittsburgh, University of Portsmouth, Princeton University, the United
States Naval Observatory, and the University of Washington.

This research has made use of the SIMBAD database, operated at CDS, Strasbourg, 
France, and of the NASA/IPAC Extragalactic Database (NED), which is operated by 
the Jet Propulsion Laboratory, California Institute of Technology, under contract 
with the National Aeronautics and Space Administration.

PyRAF and TABLES are products of the Space Telescope Science Institute, 
which is operated by AURA for NASA. 

The Digitized Sky Survey was produced at the Space Telescope Science Institute 
under U.S. Government grant NAG W-2166. The images of these surveys are based 
on photographic data obtained using the Oschin Schmidt Telescope on Palomar 
Mountain and the UK Schmidt Telescope. The plates were processed into the 
present compressed digital form with the permission of these institutions.

\end{acknowledgements}

\clearpage
 \renewcommand{\thesection}{A\arabic{section}}
 \setcounter{section}{0}  
\section*{          Appendix:     Additional Interesting Fields         }
\begin{figure} 
\centerline{\includegraphics[width=.99\columnwidth]{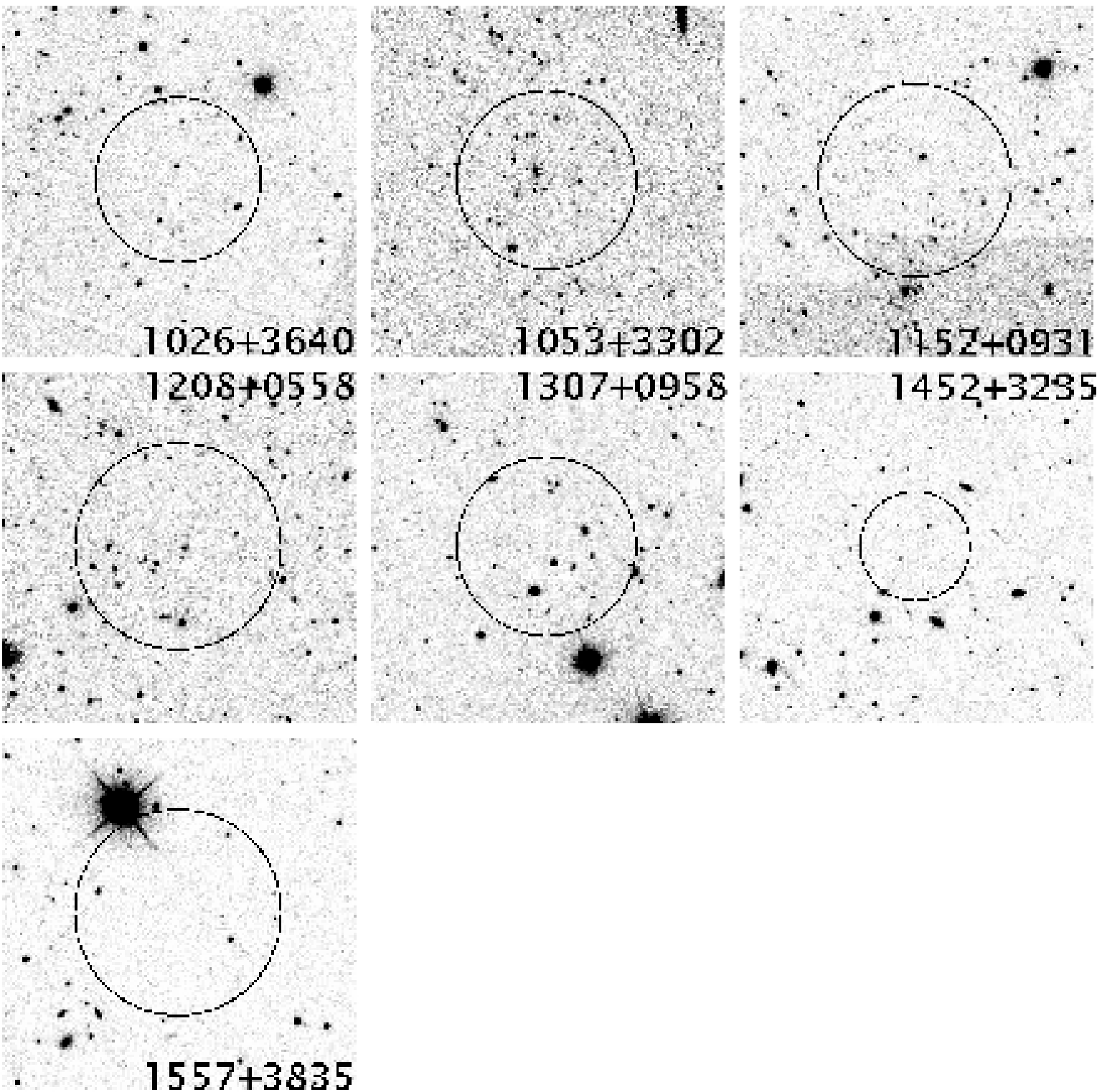}} 
\caption{SDSS composite $g,r,i$ images of candidate fields that did not survive
the cluster-detecting stage of our algorithm. The orientation and scale are as in Fig.~1.
The brightest object in any of the $4$ $p.e.$ error circles is $g = 19.35$ mag; for most fields the brightest object is $g > 21.0$. 1RXS J145234.9$+$323536 was identified as a
candidate INS by \citet{rutledge03}, but no X-ray source was detected by their 
follow-up {\it Chandra} observations.}  
\label{clusters} \end{figure}

In addition to the candidate fields listed in Table~1 and discussed in \S4, we 
present a number of interesting X-ray fields identified in the process of 
developing the algorithm described above. These fields can be divided into two 
groups. The first seven are candidate faint optical clusters, which failed 
the last step in our algorithm (see Fig.~\ref{clusters}). The other group 
is of six fields that barely fail the final version of the algorithm (all 
but one were correlated with the cluster catalog without being eliminated, 
however; see Fig.~\ref{interesting}). While this is not a complete list 
of either potential new RASS/SDSS clusters, or of interesting fields not 
quite good enough to make our final list, it does provide a sense of 
the properties of X-ray fields considered borderline optically blank by 
our algorithm.

\begin{table}[h]
\small
\begin{center}
\begin{tabular}{lcccc|cc}
\hline
\multicolumn{5}{c|}{{\it ROSAT}} & \multicolumn{2}{c}{SDSS}\\
\multicolumn{1}{c}{Source name} & & Count rate & Detection & Exp. & $g$ & Min. \\
\multicolumn{1}{c}{1RXS J} & $1\sigma$ & $10^{-2}$ cps & likelihood & s & mag & log $(f_x/f_g)$ \\
\hline
\hline
102659.6$+$364039 & $12\asec$ & $6.1\pm1.6$ & 26 & 371 & $20.43$ & $1.3$ \\
105352.5$+$330255 & $13\asec$ & $3.0\pm1.1$ & 15 & 384 & $21.05$ & $1.3$ \\
115247.0$+$093118 & $14\asec$ & $2.8\pm1.1$ & 13 & 433 & $21.51$ & $1.4$ \\
120844.3$+$055839 & $15\asec$ & $2.9\pm1.2$ & 11 & 410 & $21.50$ & $1.4$ \\
130723.7$+$095801 & $13\asec$ & $14.2\pm2.6$& 47 & 303 & $19.35$ & $1.3$ \\
145234.9$+$323536\tablenotemark{a} & $8\asec$  & $8.0\pm1.3$  & 74 & 614 & $21.52$ & $1.9$ \\
155705.0$+$383509 & $15\asec$ & $3.0\pm1.4$ & 9  & 272 & $21.12$ & $1.3$ \\
\hline
105648.6$+$413833\tablenotemark{b} & $6\asec$  & $1.6\pm0.7$ & 7  & 413 & $22.40$ & $1.5$ \\
140654.5$+$525316 & $12\asec$ & $1.3\pm0.6$ & 10 & 617 & $22.50$ & $1.5$ \\
141944.5$+$113222 & $6\asec$  & $2.2\pm1.1$ & 7  & 288 & $24.70$ & $2.6$ \\
142423.3$-$020201 & $6\asec$  & $2.6\pm1.1$ & 8  & 322 & $22.08$ & $1.6$ \\
162526.9$+$455750\tablenotemark{b} & $6\asec$  & $1.4\pm0.6$ & 7  & 572 & $22.26$ & $1.4$ \\
205334.0$-$063617\tablenotemark{b,c} & $6\asec$  & $2.1\pm0.8$ & 9  & 434 & $23.16$ & $1.9$ \\
\hline
\end{tabular}
\tablenotetext{a}{Identified as a candidate INS by \citet{rutledge03}. No source detected by {\it Chandra} \citep{rutledge03}.}
\tablenotetext{b}{Cataloged X-ray positional error appears to be an underestimate}
\tablenotetext{c}{Outside of optical cluster catalog footprint}
\caption{X-ray and optical data for seven cluster candidates. Six other interesting fields are listed below the horizontal line.}
\end{center}
\end{table}

Table~2 includes the main {\ROSAT} parameters for these RASS sources, and the 
$g$ magnitude of the {\it brightest} SDSS object within the $4$ $p.e.$ RASS 
error circle and the corresponding {\it minimum} log~$(f_x/f_g)$ for the
optical counterpart. Below we give additional information about a number of 
these fields.

\section{Candidate clusters with other possible optical counterparts}
In this section we list candidate clusters identified by our program where
another plausible optical counterpart to the X--ray source is also present 
(see \S3 for a discussion of the cluster identification stage of our algorithm).

\begin{itemize} 
\item {\bf 1RXS J102659.6$+$364039} Two potential optical counterparts to this X-ray 
source are cataloged by \citet[][]{zickgraf03}. However, both are too faint (B $> 20$)
to be unambiguously identified by Zickgraf et al.\ as the X-ray source. To
our knowledge no spectrum of either of these potential counterparts has
been taken.

\item {\bf 1RXS J130723.7$+$095801} Three optical objects within $1\amin$ of 
this X-ray source are cataloged by \citet[][]{zickgraf03}. Again, however, these 
are too faint (B $> 19$) to be unambiguously identified by Zickgraf et al.\ as 
the X-ray source; no spectrum of these potential counterparts exist to our
knowledge. 

\item {\bf 1RXS J155705.0$+$383509} This is a source with a bright ($g = 15.10$)
star near the edge of its $4$ $p.e.$ error circle. We obtained a spectrum for
this star with the APO 3.5-m telescope; it appears to be a late G/early K star
with no emission. G/K stars typically have (log) flux ratios between $-4.3$
and $-1.5$ \citep[][]{stocke91}, while this star has log~$(f_x/f_g) = -1.11$,
and is therefore unlikely to be the X-ray source.
\end{itemize}

\section{Other interesting fields}
Three of these six fields meet all of our selection criteria (1RXS J105648.6$+$413833,
1RXS 162526.9$+$455750, and 1RXS J205334.0$-$063617). However, the RASS images of
these fields, along with their count rates and exposure times, suggest that their
cataloged positional error of $6\asec$ is an underestimate. They cannot therefore 
be considered among our best INS candidates. The other three fields were identified in 
preliminary work as possibly hosting interesting X-ray sources. Below we provide
additional information about one of the $6\asec$ fields, and we describe why the
three ``early'' fields were eliminated but remain interesting.

\begin{itemize}
\item {\bf 1RXS J105648.6$+$413833} A known $g= 19.86$ quasar, 
QORG J105651.3$+$413809, is $39\asec$ ($6.4 \times$ the quoted RASS 
positional error) from the X-ray position \citep{flesch04}; it is also the
radio source FIRST J105651.2+413809. While this quasar has 
log~$(f_x/f_g) = 0.53$, within the range for AGN \citep{stocke91}, such a 
large positional offset relative to the quoted positional error means this 
quasar is unlikely to be the RASS source, unless the quoted X-ray positional 
error is underestimated.

\begin{figure}[ht] 
\centerline{\includegraphics[width=.99\columnwidth]{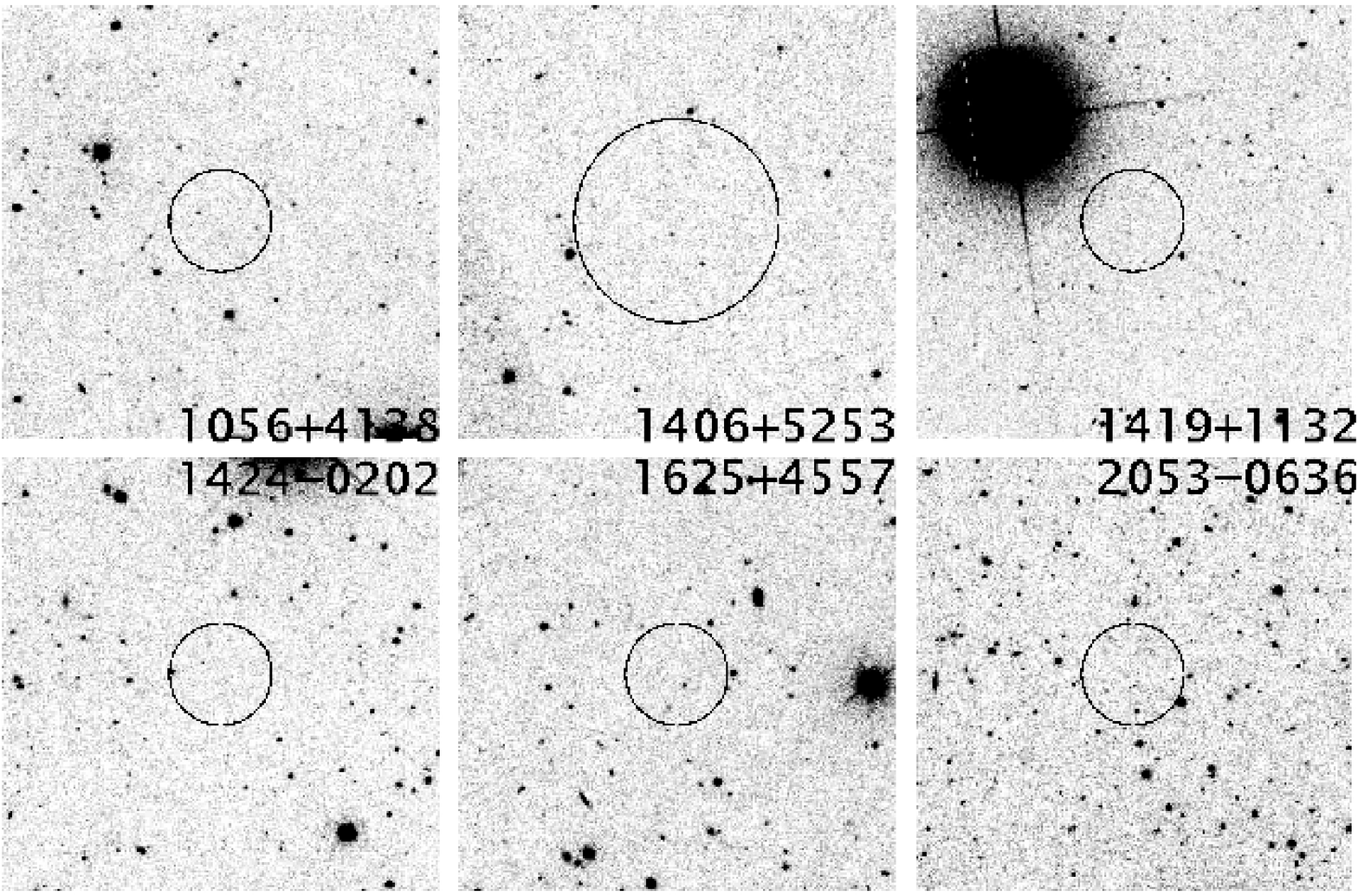}}
\caption{SDSS composite $g,r,i$ images of six fields that did not survive our 
final version of the algorithm but may host interesting X-ray emitters. The orientation
and scale are as in the previous figures. 
The brightest object in any of the $4$ $p.e.$ error circles is $g = 22.08$ mag.}
\label{interesting} \end{figure}

A {\it GALEX} \citep{galex1, galex2} source with a near-ultraviolet magnitude
of $21.85\pm0.29$ is positionally coincident ($1\asec$) with a $g= 22.67$ SDSS 
source within the $4$ $p.e.$ error circle. The nature of this object,  
SDSS J105649.58$+$413837, is difficult to determine from its SDSS photometry because
of its faintness ($u = 23.9$, $g = 22.7$) and resulting uncertainties in its 
optical colors.

\item {\bf 1RXS J140654.5$+$525316} This field was eliminated because of 
a $g = 21.70$, log $(f_x/f_g) = 1.16$ object offset from the RASS source 
by between $3$ and $4 \times$ the quoted positional error. However, this 
object's photometry is suspect and its colors are inconsistent 
($u-g = 3.5 \pm 2.5$) with that of a typical AGN.

\item {\bf 1RXS J141944.5$+$113222} This field was eliminated when the X-ray 
images of our fields were examined: it cannot be completely ruled out 
that this X-ray source and its neighbor, the bright star 
1RXS J141949.0$+$113619, are actually the same source. While it meets 
all of our algorithm's other criteria, we therefore include it in this 
list rather than among our INS candidates.

\item {\bf 1RXS J142423.3$-$020201} This field was eliminated by our algorithm
because of the presence of a $g = 20.65$ object with log $(f_x/f_g) = 1.05$ 
about $4$ $p.e.$ from the RASS position. We obtained several spectra of this object 
with the $3.5$-m telescope at APO. These spectra indicate that the object is 
most likely an ordinary G star with no signs of emission, and that it is 
therefore unlikely to be the X-ray source. The next brightest SDSS object within 
the $4$ $p.e.$ error circle is $g = 22.08$, so that the counterpart to 
the X-ray source would then have log~$(f_x/f_g) \geq 1.6$.
\end{itemize}

\end{document}